\documentstyle[epsf]{aipproc}

\newif\ifpreprint
\preprinttrue       % comment line for real proceedings

\begin{document}

\newcommand{\1}{\'{\i}}

\newcommand{\ugz }{\mbox{$\gamma$               }}
\newcommand{\unu }{\mbox{$\nu$                  }}
\newcommand{\unub}{\mbox{$\bar{\nu}$            }}
\newcommand{\upiz}{\mbox{$\pi ^{0}$             }}
\newcommand{\upin}{\mbox{$\pi \, ^{-}$          }}
\newcommand{\upip}{\mbox{$\pi \, ^{+}$          }}
\newcommand{\ulz }{\mbox{$\Lambda ^{0}$         }}
\newcommand{\ulbz}{\mbox{$\bar{\Lambda} ^{0}$   }}
\newcommand{\usp }{\mbox{$\Sigma \, ^{+}$       }}
\newcommand{\usz }{\mbox{$\Sigma \, ^{0}$       }}
\newcommand{\usn }{\mbox{$\Sigma \, ^{-}$       }}
\newcommand{\usbn}{\mbox{$\bar{\Sigma} \, ^{-}$ }}
\newcommand{\ucz }{\mbox{$\Xi \, ^{0}$          }}
\newcommand{\ucn }{\mbox{$\Xi \, ^{-}$          }}
\newcommand{\ucbp}{\mbox{$\bar{\Xi} \, ^{+}$    }}
\newcommand{\uon }{\mbox{$\Omega^{-}$           }}
\newcommand{\ukp }{\mbox{$K^{+}$                }}
\newcommand{\ukn }{\mbox{$K^{-}$                }}
\newcommand{\uks }{\mbox{$K^{0}_{s}$            }}
\newcommand{\udpp}{\mbox{$\Delta^{++}$          }}
\newcommand{\ulcp}{\mbox{$\Lambda_{c}^{+}$      }}
\newcommand{\uccp}{\mbox{$\Xi_{c}^{+}$          }}
\newcommand{\udz}{\mbox{$D^{0}$          }}
\newcommand{\udp}{\mbox{$D^{+}$          }}
\newcommand{\udn}{\mbox{$D^{-}$          }}

\newcommand{\uspg    }{\mbox{$\usp  \rightarrow p \, \ugz$              }}
\newcommand{\usbpbg  }{\mbox{$\usbn \rightarrow \bar{p} \, \ugz$        }}
\newcommand{\usppiz  }{\mbox{$\usp  \rightarrow p \, \upiz$             }}
\newcommand{\usbpbpiz}{\mbox{$\usbn \rightarrow \bar{p} \, \upiz$       }}
\newcommand{\uszl    }{\mbox{$\Sigma \Lambda$                           }}
\newcommand{\uszlg   }{\mbox{$\usz  \rightarrow \ulz \, \ugz $          }}
\newcommand{\ulppiz  }{\mbox{$\ulz  \rightarrow p \, \upiz $            }}
\newcommand{\ucsg    }{\mbox{$\ucn  \rightarrow \usn  \ugz $            }}
\newcommand{\ucnlpin }{\mbox{$\ucn  \rightarrow \ulz  \upin $           }}
\newcommand{\uocg    }{\mbox{$\uon  \rightarrow \ucn  \ugz $            }}

\newcommand{\ukppipnunub}{\mbox{$\ukp \rightarrow \upip \, \unu \, \unub$ }}
\newcommand{\ukppippiz  }{\mbox{$\ukp \rightarrow \pi^{+} \, \upiz$       }}
\newcommand{\uknpinpiz  }{\mbox{$\ukn \rightarrow \pi^{-} \, \upiz$       }}
\newcommand{\ukspinpip  }{\mbox{$\uks \rightarrow \upin \, \upip$         }}

\newcommand{\upizgg     }{\mbox{$\upiz \rightarrow \ugz \, \ugz $         }}

\newcommand{\uccppknpip }{\mbox{$\uccp \rightarrow p \, \ukn \, \upip$    }}
\newcommand{\ulcppknpip }{\mbox{$\ulcp \rightarrow p \, \ukn \, \upip$    }}
\newcommand{\udzknpip   }{\mbox{$\udz  \rightarrow \ukn \, \upip$         }}
\newcommand{\udpknpippin}{\mbox{$\udp  \rightarrow \ukn \, \upin \, \upip$}}

\newcommand{\upt}{\mbox{$p_{\perp}$ }}                       % production pt
\newcommand{\uxf}{\mbox{$x_{F}$     }}                       % production xf

\ifpreprint \rightline{\bf UASLP--IF--00--01} \fi
\title{SELEX\ifpreprint\thanks{Talk given at the
Session honoring Leon Lederman at the
VII Mexican Workshop on Particles and Fields, M\'erida, M\'exico,
November 10-17, 1999. Proceedings to be published by AIP.}\fi}
\author{Antonio Morelos }
\address{Instituto de F\'{\i}sica,
Universidad Aut\'onoma de San Luis Potos\'{\i}, \\
\'Alvaro Obreg\'on 64, Zona Centro, 78000 San Luis Potos\'{\i}, M\'exico\\
morelos@ifisica.uaslp.mx }
\maketitle
\begin{abstract}
A summary of the path which lead to a high energy physics group
at Instituto de F\1sica de la Universidad Aut\'onoma de San Luis Potos\'{\i}
is presented. 
This group is the result of the initial push made by Leon Lederman at the 
beginning of the 80's. 
\end{abstract}

By mid 80's I started as a graduate student at 
experiment E761-Fermilab. 
The first experiment where San Luis Potos\'{\i} participated is SELEX, 
experiment dedicated to charm baryon and hyperon studies. 
In parallel with SELEX another professor
joined San Luis Potos\1, together with him the group enters into a new 
challenge in experiment CKM-Fermilab including Ring Imaging Cherenkov
technology.

\section*{The history: E761} 
Experiment E761, ``An Electroweak Enigma: Hyperon Radiative Decays'' 
\cite{proposal:761} was my first encounter with experimental high energy 
physics. 

I got experience on silicon micro strip and wire chamber detectors, 
magnet spectrometers for momentum precision measurements, 
data acquisition and trigger systems based on NIM and CAMAC, 
reconstruction and analysis packages, and 
performed high precision physics measurements of 
the \usp magnetic moment and of the \usp and \usbn production polarization
\cite{am:ptxfspol,am:mmssb,am:ssbpol}.

This experiment now represents the starting push for the particle physics
group at San Luis Potos\1.

\ifpreprint\goodbreak\else\newpage\fi
\section*{The present: SELEX}

There are several aspects to talk about SELEX, recent hyperon and charm 
physics results, I only highlight a few of them. 

Also prior to talk about physics I should mention that E781 or SELEX is the 
first experiment where San Luis Potos\1 is a formal collaborating institution. 
Two students have gotten their master degree thesis related to SELEX, 
Ricardo L\'opez Fern\'andez working in the RICH group and analyzing the beam 
composition using the RICH, and Galileo Dom\'{\i}nguez Zacar\'{\i}as working 
with the \uks sample looking into $\pi$ asymmetry \cite{uaslp:trlf,uaslp:tgdz}.
Personally I worked in the smart crate controller in the CAMAC setup and 
trigger installation. 
An impact at San Luis Potos\1 also happen in relation to this collaboration: 
J\"urgen Engelfried, member of SELEX and previously WA89 at CERN, 
accepted to work as professor at San Luis 
Potos\1, his expertise brings more life and conforms the local group and
opens more physics opportunities as I'll discuss in the section 
``The near future''.

SELEX is a new fixed target experiment designed to enhance charm - strange 
baryon over meson data. The data taking lasted from summer 1996 to fall 1997. 
It includes a tagged hadron beam on $\pi$, $\Sigma$ and proton using a TRD 
detector; a micro-vertex detector, and particle id using TRD, 
lead glass, and RICH detectors; all these detectors distributed among
three magnet spectrometers \cite{proposal:781}.

SELEX was designed to be a high \uxf charm baryon spectrometer, and in fact 
this can be appreciated by the high acceptance at \uxf $>$ 0.5. In 
the three modes, \ulcppknpip, \udzknpip, and \udpknpippin, 
the acceptance is grater than 6 \%, and identical for particle and antiparticle
decays. The control on the acceptance gives
the opportunity to study charm baryon production as a function of \uxf with 
very good precision for challenging theoretical models and other 
experiments \cite{talk:fernanda,fg:ucla,jr:vancuver}.

Charm baryon and meson lifetime measurements are also under control in the 
experiment. SELEX was designed with a trigger to enhance all charm baryon 
production and decay modes, as a result it has the ability to look for unseen
decay modes, right now SELEX is reporting the first observation of Cabibbo 
suppressed \uccppknpip decay \cite{selex:cabibbo}.

The fact of having a \usn beam to study charm production also provides by 
its own nature the tool to study properties of the hyperon itself. At present
there are preliminary reports on \usn radius and total cross section of
\usn beam on different target material 
\cite{uwe:total,mam:inelastic,vk:ados,ivo:hyperon}.

\section*{The near future: CKM \& Instrumentation}

On April 26 1996, Fermilab invited physicists for a workshop towards the
use of a 120 GeV/c proton beam on collider and fixed target mode. All 
mexican groups had the opportunity for joining this activity. 
The event
really represented a great opportunity for young mexican groups since the 
enterprises are small in size and with a lot of opportunities to start at 
the zero point of the design and construction for leading a project or
subproject. Of all the mexican existing groups only San Luis Potos\1, so far, 
has taken the challenge to participate actively in one of these experiments.

CKM ``Charged Kaons at the Main Injector'', a proposal for a Precision 
Measurement of the Decay \ukppipnunub and Other Rare \ukp Processes at 
Fermilab 
Using the Main Injector, is one of the experiments which were born after 
the April 96 workshop \cite{proposal:ckm}. 

The experiment will measure the branching ratio of the  
ultra-rare charged kaon decay \ukppipnunub by observing a large sample of  
those decays with small background.  The physics goal is to measure the 
magnitude of the Cabibbo, Kobayashi, Maskawa matrix element $V_{td}$ with a 
statistical precision of about 5\% based upon a $\sim$~100 event sample with 
total backgrounds of  less than 10 events.  This decay mode is known to be 
theoretically clean. The only significant theoretical uncertainty in the 
calculation of this branching ratio is due to the charm contribution.  A 
10\% measurement of the branching ratio will yield a 10\% total uncertainty 
on the magnitude of $V_{td}$. 

In this experiment IF-UASLP is in charge for testing
parts and the whole design of two Ring Imaging Cherenkov Counters, RICH's
\cite{plan:erik}.
Experience on this technology came from the participation of J\"urgen 
Engelfried on the design, construction, operation, and analysis of
two previous RICH's, one in WA89 and another in SELEX. Also, in SELEX, 
Ricardo L\'opez Fern\'andez, graduate student
from IF-UASLP, worked on and used the RICH as part of his M.Sc.\ thesis.
 
CKM marks the near future experimental enterprise we are working on at 
IF-UASLP. We are initiating high energy physics instrumentation at IF-UASLP, 
aimed right now at the RICH technology applied to the CKM experiment.

\section*{HEP: a group at IF-UASLP}

Experimental high energy physics evolves with projects and this aspect is
also reflected on the IF-UASLP group which has seen the pass of experiments
E761, WA89, E781 and now CKM. 
Presently, the group is creating a high energy  instrumentation 
laboratory towards detector research and development. 
The basic idea of the laboratory is to target user defined detectors at world 
wide experiments, right now we have the RICH design and testing for CKM.
In 1999, IF-UASLP just hired Ruben Flores Mendieta
to strength particle physics theory and phenomenology research.
Looking backwards from the beginning of the 80's to the end of the 90's a 
spawn of close to 20 years has happened for initiating an experimental group at
San Luis Potos\1, that is a positive result from an initial kick.

\section*{Acknowledgement}

This work was partly financed by IF-UASLP and CONACyT.

\end{document}